\documentclass[pra,aps,superscriptaddress,twocolumn,color]{revtex4-1}

\usepackage{amsmath}
\usepackage{amsfonts}
\usepackage{bm}
\usepackage{graphicx}
\usepackage{array,color}

\begin{document}
\title{Hydrodynamic generation of skyrmions in a two-component Bose-Einstein
condensate}

\author{Kyoshiro Sakaguchi}
\affiliation{Department of Engineering Science, University of
Electro-Communications, Tokyo 182-8585, Japan}

\author{Keisuke Jimbo}
\affiliation{Department of Engineering Science, University of
Electro-Communications, Tokyo 182-8585, Japan}

\author{Hiroki Saito}
\affiliation {Department of Engineering Science, University of
Electro-Communications, Tokyo 182-8585, Japan}

\date{\today}

\begin{abstract}
When an obstacle is moved in a superfluid faster than a critical velocity,
quantized vortices are generated behind the obstacle.
Here we propose a method to create more complicated topological
excitations, three-dimensional skyrmions, behind a moving obstacle.
We numerically show that, in a two-component Bose-Einstein condensate,
component-dependent obstacle potentials can generate skyrmions in the wake,
made up of quantized vortex rings in different components that are linked
with each other.
The lifetime of generated skyrmions can be prolonged by a guiding potential,
which enables the formation of a skyrmion train.
\end{abstract}

\maketitle

\section{Introduction}
\label{s:intro}

Skyrmions are particle-like topological excitations of fields.
They were originally proposed to describe mesons and baryons in
nuclear physics~\cite{Skyrme}, and were later applied to other physical
systems, such as quantum Hall systems~\cite{Lee,Brey,Schmeller}, magnetic
materials~\cite{Bogdanov}, liquid crystals~\cite{Bogdanov2}, and
multicomponent Bose-Einstein condensates (BECs)~\cite{Ruo, Khawaja}.
In these systems, skyrmions can be generated by various means.
For example, in a magnetic film, two-dimensional skyrmions can be created by
applying a magnetic field perpendicular to the surface~\cite{Yu}.
Using the tip of a scanning tunneling microscope, magnetic skyrmions can be
written or deleted in a controlled manner~\cite{Romming}.

Here we focus on the generation schemes of skyrmions in a multicomponent
BEC.
Experimentally, two-dimensional skyrmions have been created by Raman
transitions~\cite{Leslie} and by magnetic-field induced spin
rotations~\cite{Choi}.
Three-dimensional skyrmions were recently realized in a spin-1 BEC by spin
rotation with a controlled magnetic field~\cite{Lee2}.
Theoretically, several schemes to generate skyrmions in multicomponent BECs
have been proposed: Rabi transitions with topological phases~\cite{Ruo},
spin rotation by a fictitious or real magnetic
field~\cite{Khawaja,Tiurev,Zamora,Luo},
capillary instability~\cite{Sasaki},
spin-orbit coupling~\cite{Kawakami, Zhang},
decay of domain walls~\cite{Nitta},
and light-matter coupling~\cite{Parmee}.
The stability and dynamics of skyrmions have also been
studied~\cite{Khawaja2, Battye, Zhang2, Zhai, Savage, Ruo2, Wuster, Herbut,
  Tokuno, Price, Kaneda}.
In this paper, we propose an alternative scheme to generate
three-dimensional skyrmions in a two-component BEC: hydrodynamic generation
of skyrmions behind an obstacle moving in the BEC.
This scheme is reminiscent of that in Ref.~\cite{Kleckner}, in which trefoil
and linked vortices are created in water using hydrofoils with special
shapes.
Such generation schemes of topological excitations explore an
interdisciplinary field of hydrodynamics and topology, paving the way for
understanding such phenomena as the topological excitations in quantum
turbulence~\cite{Cooper}.

A quantized vortex is the simplest topological excitation in a BEC, and
can be generated by an external potential moving in a BEC~\cite{Frisch,
Jackson, Nore}.
Such hydrodynamic generation of quantized vortices has been realized in
experiments, where pairs of vortices and antivortices (vortex dipoles) were
created behind Gaussian laser beams swept through
BECs~\cite{Inouye, Neely, Kwon, Kwon2}.
The successive generation of vortices by a moving obstacle potential forms a
periodic pattern, which is a quantum analogue of the B\'enard-von K\'arm\'an
vortex street~\cite{SasakiL, Kwon3}.
In a spinor or two-component BEC, a moving obstacle potential can be used to
generate half-quantum vortices~\cite{Seo, Seo2}.
Quantized vortices have also been observed in exciton-polariton superfluids
flowing around obstacle potentials~\cite{Nardin}.
Thus, so far, the hydrodynamic generation of topological excitations in
superfluids has been restricted to quantized or half-quantized vortices,
which are essentially topological structures in two dimensions.

The aim of the present study is to generate more intriguing topological
excitations, three-dimensional skyrmions, by an obstacle potential moving in
a miscible two-component BEC.
We propose a special configuration of external potentials that depend on the
components, and move them in the two-component BEC.
We will show that quantized vortex rings generated in different components
are linked with each other.
Such a structure is characterized by a nonzero integer winding number and
is regarded as a skyrmion.
As in the case of quantized-vortex generation, successive generation of
skyrmions is possible in this method.
When the generated skyrmions are kept stable, they form a skyrmion train,
just like a vortex street behind an obstacle.

The remainder of the paper is organized as follows.
Section~\ref{s:formulation} reviews structures of skyrmions in a
two-component BEC and defines a topological structure to be created
hydrodynamically.
Section~\ref{s:uniform} proposes potential configurations for creating
skyrmions and numerically demonstrates skyrmion generation in an ideal
uniform system.
Section~\ref{s:trap} studies a realistic system confined in a harmonic
potential.
Section~\ref{s:conc} provides conclusions to this study.

\section{Skyrmion in a two-component Bose-Einstein condensate}
\label{s:formulation}

We consider a two-component BEC at zero temperature, described by the
macroscopic wave functions $\psi_1(\bm{r}, t)$ and $\psi_2(\bm{r}, t)$ in
the mean-field approximation.
The two-component wave functions can generally be written as
\begin{equation}
\boldsymbol\Psi(\bm{r}) =
  \left( \begin{array}{c} \psi_1(\bm{r}) \\ \psi_2(\bm{r}) \end{array}
  \right) = \sqrt{\rho(\bm{r})} \left( \begin{array}{c} \zeta_1(\bm{r})
      \\ \zeta_2(\bm{r}) \end{array} \right),
\end{equation}
where $\rho = |\psi_1|^2 + |\psi_2|^2$ is the total density and
$|\zeta_1|^2 + |\zeta_2|^2 = 1$.
(We assume $\rho \neq 0$ to avoid discontinuities in $\bm{\zeta}$.)
The vector $(\zeta_1, \zeta_2)$ can be regarded as a state of
pseudospin-1/2, which has the SU(2) manifold.
A skyrmion in a two-component BEC is defined as a state in which the
physical space $\bm{r}$ is continuously mapped onto the SU(2) manifold in a
topologically nontrivial manner.
In this mapping, $(\zeta_1, \zeta_2)$ must go to a common state at infinity
($r \rightarrow \infty$).
Mathematically, the topology of this map is represented by the third
homotopy group $\pi_3(SU(2)) = \mathbb{Z}$ and the skyrmion is
characterized by an integer winding number.

A simple expression of a skyrmion is given by
\begin{equation} \label{skyrmion}
  \boldsymbol\Psi(\bm{r}) = \sqrt{\rho}
  e^{-i \chi(r) \bm{\sigma} \cdot \bm{n}}
  \left( \begin{array}{c} 1 \\ 0 \end{array} \right)
  = \sqrt{\rho} \left( \begin{array}{c} \cos\chi - i \sin\chi \cos\theta \\
  -i \sin\chi \sin\theta e^{i\phi} \end{array} \right),
\end{equation}
where $\bm{\sigma} = (\sigma_x, \sigma_y, \sigma_z)$ is the vector of the
Pauli matrices and $\bm{n} = \bm{r} / r = (\sin\theta \cos\phi, \sin\theta
\sin\phi, \cos\theta)$ is a unit vector pointing in the radial direction
with polar coordinates $(r, \theta, \phi)$.
The continuous function $\chi(r)$ must be integer multiples of $\pi$ at $r =
0$ and $r = \infty$ in order that the state does not depend on $\theta$ and
$\phi$ at the origin and infinity.
If we impose the boundary condition $\chi(0) = 0$ and $\chi(\infty) = \pi$,
$\exp(-i\chi \bm{\sigma} \cdot \bm{n})$ covers whole elements of SU(2).
In fact, Eq.~(\ref{skyrmion}) runs over whole spin states.
It is apparent from the real and imaginary parts of each component in
Eq.~(\ref{skyrmion}) that the manifold corresponds to the surface of a unit
sphere in four-dimensions with an area $2\pi^2$.
The winding number is defined by the number of times that the sphere is
wrapped around:
\begin{equation} \label{W0}
  W = \frac{1}{2\pi^2} \int_0^{\chi(\infty)} d\chi \int_0^{\pi} d\theta
  \int_0^{2\pi} d\phi \sin^2 \chi \sin \theta.
\end{equation}
When $\chi(r)$ changes from $\chi(0) = 0$ to $\chi(\infty) = \ell \pi$ with
an integer $\ell$, $W = \ell$.

The variables $\chi$, $\theta$, and $\phi$ in Eq.~(\ref{skyrmion}) can be
regarded as functions of the Cartesian coordinates $\bm{r} = (x, y, z)$.
Let us allow continuous deformation of these functions, which we denote as
$\alpha(\bm{r})$, $\beta(\bm{r})$, and $\gamma(\bm{r})$.
A general form of the wave function can then be expressed as
\begin{equation}
  \Psi(\bm{r}) = \sqrt{\rho(\bm{r})}
  \left( \begin{array}{c} \cos\alpha(\bm{r}) - i \sin\alpha(\bm{r})
    \cos\beta(\bm{r}) \\
  -i \sin\alpha(\bm{r}) \sin\beta(\bm{r}) e^{i\gamma(\bm{r})} \end{array}
  \right).
\end{equation}
Changing the integration variables from those in Eq.~(\ref{W0}) to the
Cartesian coordinates, we obtain a general form of the winding number as
\begin{equation} \label{W}
  W = \frac{1}{2\pi^2} \int d\bm{r} \sin^2 \alpha(\bm{r}) \sin \beta(\bm{r})
  {\rm det} \left(
  \frac{\partial(\alpha, \beta, \gamma)}{\partial(x, y, z)} \right),
\end{equation}
where ${\rm det}(\cdots)$ is the Jacobian.
The winding number in Eq.~(\ref{W}) is an integer as long as the wave
function is continuous.

The macroscopic wave functions obey the coupled Gross-Pitaevskii (GP)
equations given by
\begin{subequations} \label{GP}
\begin{eqnarray}
i\hbar \frac{\partial \psi_1}{\partial t} & = & \left( -\frac{\hbar^2}{2m_1}
\nabla^2 + V_1 + g_{11} |\psi_1|^2 + g_{12} |\psi_2|^2 \right) \psi_1,
\nonumber \\
\\
i\hbar \frac{\partial \psi_2}{\partial t} & = & \left( -\frac{\hbar^2}{2m_2}
\nabla^2 + V_2 + g_{22} |\psi_2|^2 + g_{12} |\psi_1|^2 \right) \psi_2,
\nonumber \\
\end{eqnarray}
\end{subequations}
where $m_j$ is the atomic mass of component $j$, $V_j(\bm{r}, t)$ is the
external potential for component $j$, and $g_{jj'} = 2\pi\hbar^2 a_{jj'} /
m_{jj'}$ is the interaction coefficient with $a_{jj'}$ and $m_{jj'}$ being
the $s$-wave scattering length and reduced mass between components $j$ and
$j'$.
Here, for simplicity, we assume that the atoms in both components have the
same mass $m \equiv m_1 = m_2$ and the same intracomponent interaction $g
\equiv g_{11} = g_{22}$.
In this section, we consider the case in which $V_1 = V_2 = 0$.
In numerical simulations, the real-time and imaginary-time evolutions of
Eq.~(\ref{GP}) are solved using the pseudospectral method~\cite{recipe}.
In the results shown in this section and in Sec.~\ref{s:uniform}, length,
time, energy, and density are normalized by $\hbar / (2m g n_0)^{1/2}$,
$\hbar / (g n_0)$, $g n_0$, and $n_0$, respectively, where $n_0$ is the
uniform density far from the skyrmion.

\begin{figure}[tb]
\includegraphics[width=8.5cm]{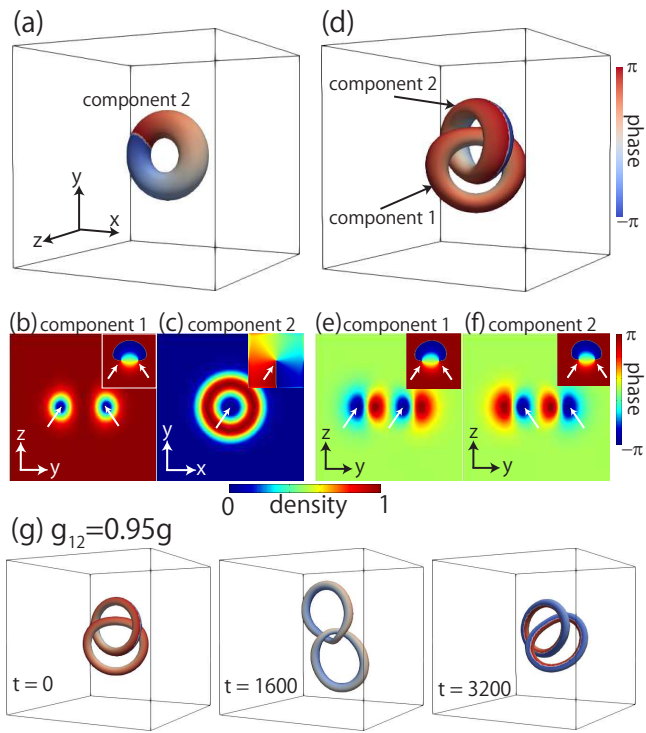}
\caption{
  (a)-(c) Skyrmion state obtained by imaginary-time propagation of the GP
  equation starting from Eq.~(\ref{skyrmion}), where $g_{12} = g$.
  (a) Isodensity surface of component 2 ($|\psi_2|^2 = 0.5$).
  (b) Density and phase (inset) profiles of component 1 on the plane of
  $x = 0$.
  (c) Density and phase (inset) profiles of component 2 on the plane of
  $z = 0$.
  (d)-(f) Spin state of (a)-(c) rotated by $\pi/2$ using
  Eq.~(\ref{rotate}).
  (d) Isodensity surfaces of both components ($|\psi_1|^2 = |\psi_1|^2 =
  0.2$).
  (e), (f) Density and phase (insets) profiles of components 1 and 2 on the
  plane of $x = 0$.
  The white arrows indicate the positions of quantized vortex cores.
  (g) Time evolution of the isodensity surfaces for $g_{12} = 0.95 g$.
  The size of the boxes in (a), (d), and (g) is $128 \times 128 \times 128$
  in units of the healing length.
  See the Supplemental Material for videos of the real-time dynamics in (a),
  (d), and (g)~\cite{movies}.
}
\label{f:skyrmion}
\end{figure}
Let us consider a skyrmion structure in Eq.~(\ref{skyrmion}) with a winding
number $W = 1$.
This state consists of a quantized vortex ring in component 1 and a
toroidal-shaped component 2 with a quantized circulation along the
torus~\cite{Ruo}, as shown in Figs.~\ref{f:skyrmion}(a)-\ref{f:skyrmion}(c).
The core of the vortex ring of component 1 is filled with component 2, and
the total density is almost constant.
Numerically, this skyrmion state is obtained by a short imaginary-time
evolution (duration of $\sim 100$) of the GP equation starting from an
initial state in Eq.~(\ref{skyrmion}) with an appropriate $\chi(r)$.
Since the skyrmion state is not a local minimum of the energy, long
imaginary-time evolution eliminates the skyrmion or expands it to infinity.
After the short imaginary-time evolution, a small numerical noise is added
to break the numerically exact symmetry, followed by the real-time
evolution.
The dynamics of the skyrmion for $g = g_{12}$ shown in the video provided in
the Supplemental Material~\cite{movies} indicates that the skyrmion is
stable.
Since the vortex ring has a momentum along the axis of the ring ($z$
direction), the skyrmion moves at a constant velocity in a uniform
system~\cite{Kaneda}.
We therefore use the frame of reference moving with the skyrmion by adding a
term $i \hbar v_z \partial_z \psi_j$ to the right-hand side of
Eq.~(\ref{GP}).
The velocity $v_z$ is chosen to be 0.08 so that the skyrmion is at rest in
the real-time evolution.

In most previous studies, the structure as shown in
Figs.~\ref{f:skyrmion}(a)-\ref{f:skyrmion}(c) has been studied as a skyrmion
in a two-component BEC, which is also referred to as a
vorton~\cite{Nitta}.
Alternatively, topologically equivalent states can be obtained by rotating
the state in the pseudospin space, since the winding number $W$ is unchanged
by the global spin rotation.
Figures~\ref{f:skyrmion}(d)-\ref{f:skyrmion}(f) show a state that is
obtained by the spin rotation of the state in Figs.~\ref{f:skyrmion}(a)-(c)
by $\pi / 2$ around the $y$ axis,
\begin{equation} \label{rotate}
  \exp\left(-i \frac{\pi}{2} \frac{\sigma_y}{2}\right)
  = \frac{1}{\sqrt{2}} \left( \begin{array}{cc} 1 & -1 \\ 1 & 1 \end{array}
  \right).
\end{equation}
In this skyrmion state, both components contain vortex rings, and the two
vortex rings in components 1 and 2 link with each other~\cite{Gudnason}, as
shown in Fig.~\ref{f:skyrmion}(d).
The dynamics of this skyrmion state is shown in the video provided in the
Supplemental Material~\cite{movies}.
The linked vortices slowly revolve about the $z$ axis at a frequency of
$\sim 10^{-3}$.
This rotation can be understood from the fact that, in the original state in
Figs.~\ref{f:skyrmion}(a)-(c), the torus-shaped component 2 and the
surrounding component 1 have different chemical potentials $\mu_1 \neq
\mu_2$, resulting in an increase in the relative phase between the two
components as $(\mu_1 - \mu_2) t / \hbar$.
The relative phase also increases with the angle $\phi$ around the $z$ axis,
since the torus of component 2 has a quantized vortex, as shown in
Figs.~\ref{f:skyrmion}(a) and \ref{f:skyrmion}(c), and thus the relative
phase is $(\mu_1 - \mu_2) t / \hbar - \phi$.
Thus, the spin structure in Fig.~\ref{f:skyrmion}(d) revolves about the $z$
axis at a frequency $(\mu_1 - \mu_2) / h$, since the relative phase between
the two components before the spin rotation determines the spin structure of
Fig.~\ref{f:skyrmion}(d).
The stability of the state in Figs.~\ref{f:skyrmion}(d)-\ref{f:skyrmion}(f)
is the same as that of Figs.~\ref{f:skyrmion}(a)-\ref{f:skyrmion}(c), since
the GP equation is invariant with respect to the spin rotation for $g =
g_{12}$.

We examine the stability of the skyrmion state for $g_{12} < g$.
Figure~\ref{f:skyrmion}(g) shows the time evolution of the skyrmion state
with linked vortices for $g_{12} = 0.95g$.
The initial state is prepared by a short imaginary-time evolution of the
state in Fig.~\ref{f:skyrmion}(d) with $g_{12} = 0.95g$.
In Fig.~\ref{f:skyrmion}(g), the linked vortices are first stretched at $t =
1600$ and nearly divide into individual vortex rings, which are however kept
linked after that.
The link survives at least until $t = 3200$.
Through the time evolution, the winding number $W$ is kept to almost unity.
If the state in Fig.~\ref{f:skyrmion}(d) (prepared with $g_{12} = g$) is
used as the initial state of the time evolution for $g_{12} = 0.95g$
(without the short imaginary-time evolution), the linked vortices are
unlinked at $t \simeq 800$ (data not shown) because of the excess energy due
to the change of $g_{12}$, which was reduced by the short imaginary-time
evolution in Fig.~\ref{f:skyrmion}(g).
These results imply that there is an energy barrier to unlink the vortex
rings.
This is because there must appear a phase singularity with zero density
(U(1) vortex) at which the vortices are unlinked.

For the present purpose, i.e., hydrodynamic generation of skyrmions by a
moving obstacle, the skyrmion state as in
Figs.~\ref{f:skyrmion}(d)-\ref{f:skyrmion}(g) (linked vortices) is more
suited than that in Figs.~\ref{f:skyrmion}(a)-\ref{f:skyrmion}(c) (vorton).
The vorton state cannot be created just by moving an obstacle in component
1, because $N_2$ is conserved within the GP equation; a localized component
2 must be added to create a vorton.
On the other hand, for the linked-vortex state generation, such an extra
procedure is not needed, since both components exist from the start.
In the next section, we aim to produce the linked-vortex state by a moving
obstacle.

\section{Skyrmion generation by a moving obstacle}

\subsection{Uniform system}
\label{s:uniform}

\begin{figure}[tb]
\includegraphics[width=8.5cm]{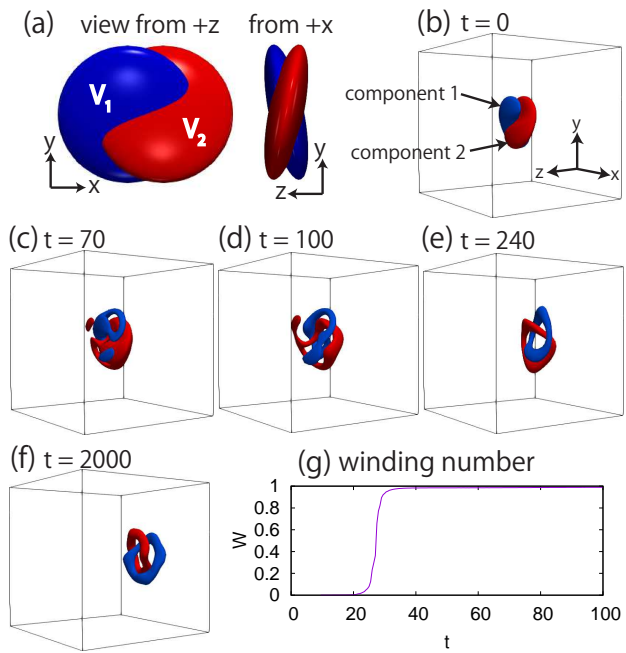}
\caption{
  (a) Isosurfaces of the obstacle potentials ($V_1 = V_2 = 0.5 V_0$) in
  Eq.~(\ref{Vj}).
  The parameters of the potential are $V_0 = 1$, $\delta_1 = -\delta_2 =
  3$, $\lambda_1 = -\lambda_2 = \pi / 16$, and $R = 14$.
  (b)-(f) Dynamics of skyrmion generation for $g_{12} = 0.98 g$.
  The potential in (a) is moved at a velocity $v_z = 0.18$, which is
  linearly ramped down and vanishes at $t = 100$.
  Isodensity surfaces of both components ($|\psi_1|^2 = |\psi_2|^2 = 0.2$)
  are shown, where the densities are smaller inside the surfaces.
  The frame of reference is moved at $v_z = 0.18$ in the $z$ direction.
  The size of the boxes in (b)-(f) is $128 \times 128 \times 128$
  in units of the healing length.
  See the Supplemental Material for a video of the dynamics in
  (b)-(f)~\cite{movies}.
  (g) Time evolution of the winding number $W$ defined in Eq.~(\ref{W}).
}
\label{f:single}
\end{figure}
In this subsection, we consider an ideal uniform system without a trapping
potential.
We use an obstacle potential that is different for the components 1 and 2,
$V_1 \neq V_2$.
Such spin dependent potentials are realized by near-resonant laser
beams~\cite{Kim, Kim2}.
Here, we propose an obstacle potential given by $(j = 1,2)$
\begin{eqnarray} \label{Vj}
  V_j(\bm{r}, t) & = & V_0(t) e^{-[(x + \delta_j)^2 + \eta^2 + 4 \zeta^2] /
    R^2}, \nonumber \\
  \eta & = & y \cos\lambda_j - z \sin\lambda_j,
  \nonumber \\
  \zeta & = & z \cos\lambda_j + y \sin\lambda_j,
\end{eqnarray}
where $V_0$ is the strength of the oblate Gaussian potential, and $\delta_j$
and $\lambda_j$ are the spin-dependent shift in the $x$ direction and
rotation angle about the $x$ axis.
The shape of this potential is depicted in Fig.~\ref{f:single}(a), where the
parameters are taken to be $V_0 = 1$, $\delta_1 = -\delta_2 = 3$, 
$\lambda_1 = -\lambda_2 = \pi / 16$, and $R = 14$.
The two oblate potentials for components 1 and 2 are tilted and shifted in
opposite directions, and they partly overlap with each other.
An important point in this potential configuration is that the circular
edges of these spheroids are linked with each other, so that they resemble
the configuration of the linked vortex rings in the skyrmion in
Fig.~\ref{f:skyrmion}(d).

In the numerical simulation of the GP equation, the ground state in the
presence of the potential in Fig.~\ref{f:single}(a) is prepared by the
imaginary-time evolution.
In the subsequent real-time evolution, the term $i\hbar v_z \partial_z
\psi_j$ is introduced in the right-hand side of the GP equation, which
corresponds to the situation in which the potential starts to move in the
$z$ direction at a velocity $v_z$.
At the same time, the strength of the potential $V_0(t)$ is linearly ramped
down from $V_0(0) = 1$ to $V_0(100) = 0$.
After $t = 100$, the potential is kept at zero.

Figures~\ref{f:single}(b)-\ref{f:single}(f) demonstrate the real-time
dynamics of the skyrmion generation behind the obstacle potential.
In these figures, the tube-like surfaces contain vortex lines of the
corresponding component filled with the other component, i.e., they indicate
half-quantum vortex lines.
First, half-quantum vortex lines emerge from the upstream potential edges
(Fig.~\ref{f:single}(c)).
After the potential vanishes ($t > 100$), the linked vortex rings remain
(Figs.~\ref{f:single}(d) and \ref{f:single}(e)), which is a skyrmion similar
to that in Fig.~\ref{f:skyrmion}(d).
The winding number $W$ increases from 0 to $\simeq 1$, as shown in
Fig.~\ref{f:skyrmion}(g).
The mechanism of the skyrmion generation is quite simple:
the vortex line for each component is generated near the circular edge of
the oblate potential, where the flow velocity exceeds the critical velocity
of vortex generation.
The skyrmion is thus formed from the linked configuration of the circular
edges of the oblate potentials.
The skyrmion survives for a long time, as shown in Fig.~\ref{f:single}(f).
The lifetime of a generated skyrmion is typically $\sim 1000$.
In Figs.~\ref{f:single}(b)-\ref{f:single}(f), the velocity of a skyrmion
moving in the $+z$ direction is slower than $v_z$, and it travels in the
$-z$ direction in the moving frame of reference.

\begin{figure}[tb]
\includegraphics[width=8.0cm]{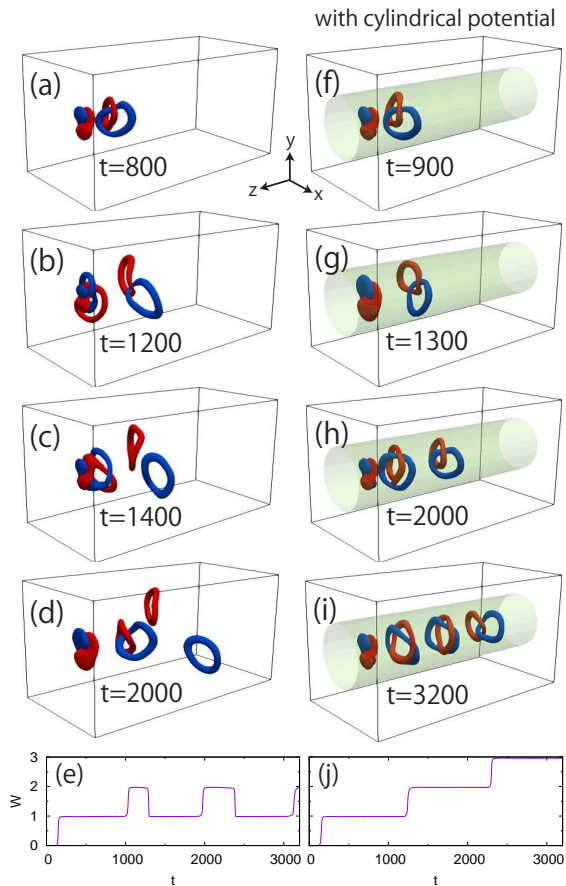}
\caption{
  Sequential generation of skyrmions for $g_{12} = 0.95 g$, $v_z = 0.16$,
  $V_0 = 1$, $\delta_1 = -\delta_2 = 0.5$, $\lambda_1 = -\lambda_2 = \pi /
  24$, and $R = 10$.
  The strength of the potential $V_0$ is constant.
  (a)-(e) Uniform system.
  (f)-(j) Cylindrical potential in Eq.~(\ref{cyl}) with $R_c = 32$ is
  added.
  In (a)-(d) and (f)-(i), isodensity surfaces of both components
  ($|\psi_1|^2 = |\psi_2|^2 = 0.2$) are shown.
  The size of the boxes is $128 \times 128 \times 256$
  in units of the healing length.
  In (f)-(i), the radius of the cylinder is shown by shading.
  See the Supplemental Material for videos of the dynamics in (a)-(d) and
  (f)-(i)~\cite{movies}.
  (e), (j) Time evolution of the winding number $W$ without and with the
  cylindrical potential.
}
\label{f:multi}
\end{figure}
Figure~\ref{f:multi} shows the sequential generation of skyrmions, where the
strength $V_0$ of the potential is kept constant.
In this case, after the skyrmion is generated behind the potential
(Fig.~\ref{f:multi}(a)), the two linked vortex rings are stretched
(Fig.~\ref{f:multi}(b)) and then detach from each other at $t \simeq 1300$
(Fig.~\ref{f:multi}(c)).
When the linked vortex rings are detached, the winding number $W$ is
decreased by $\simeq 1$.
Although the skyrmions are created sequentially with a period of $\sim
1000$, they are broken with a lifetime of $\sim 1000$, resulting in an
oscillation of $W$, as shown in Fig.~\ref{f:multi}(e).
We examined various parameters and found that the skyrmions always have
finite lifetimes, which makes it difficult to maintain three or more
skyrmions at the same time.

To enhance the stability of the created skyrmions, we introduce a
cylindrical potential,
\begin{equation} \label{cyl}
  V_{\rm cyl}(\bm{r}) = \left\{ \begin{array}{ll} 0 & (r_\perp \leq R_c) \\
    -g n_0 & (r_\perp > R_c), \end{array} \right.
\end{equation}
where $r_\perp = (x^2 + y^2)^{1/2}$.
This potential makes the density at $r_\perp > R_c$ large.
The vortices tend to be confined within the cylinder of $r_\perp \leq R_c$,
which can prevent the stretching and detaching of vortices in the transverse
direction, as shown in Figs.~\ref{f:multi}(b) and \ref{f:multi}(c).
Figures~\ref{f:multi}(f)-\ref{f:multi}(i) show the dynamics in the presence
of the cylindrical potential with $R_c = 32$.
The wave functions are normalized in such a way that the density inside the
cylinder is almost the same as that without the cylinder.
We see that the cylindrical potential serves as a guide for the skyrmions,
and the skyrmions generated behind the obstacle potential are maintained,
forming a skyrmion train.
The winding number $W$ thus increases monotonically, as shown in
Fig.~\ref{f:multi}(j).

\begin{figure}[tb]
\includegraphics[width=8.0cm]{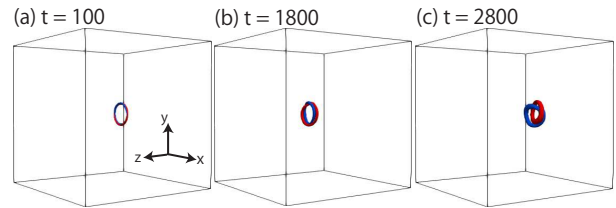}
\caption{
  Growth of a skyrmion from overlapped vortex rings for $g_{12} = 0.95 g$.
  The vortex rings are generated by the potential in Eq.~(\ref{Vj}) with
  $V_0 = 1$, $\delta_1 = -\delta_2 = 0.5$, $\lambda_1 = -\lambda_2 = \pi /
  200$, $R = 14$, and $v_z = 0.35$.
  The potential is linearly ramped down and vanishes at $t = 100$.
  Isodensity surfaces of both components ($|\psi_1|^2 = |\psi_2|^2 = 0.2$)
  are shown.
  The size of the boxes is $128 \times 128 \times 128$
  in units of the healing length.
  See the Supplemental Material for a video of the dynamics~\cite{movies}.
}
\label{f:stability}
\end{figure}
Next, we study the case of small shifts $\delta_j$ and rotation angles
$\lambda_j$ of the potentials in Eq.~(\ref{Vj}), i.e., the ellipsoidal
potentials in the two components almost overlap.
In this case, unlike Figs.~\ref{f:single} and \ref{f:multi}, the vortex lines
generated in different components almost overlap.
Figure~\ref{f:stability} shows the dynamics of vortices generated by the
potential with $\delta_1 = -\delta_2 = 0.5$, $\lambda_1 = -\lambda_2 = \pi /
200$, which are much smaller than those in Figs.~\ref{f:single}
and~\ref{f:multi}.
Since the vortex cores generated in the two components initially overlap,
they practically form a vortex ring with the U(1) vortex line, and
therefore, the thickness of the vortex core is smaller than those of
half-quantum vortex lines in a skyrmion, as shown in
Fig.~\ref{f:stability}(a).
As time elapses, the small deviation between the two vortex rings grows
(Fig.~\ref{f:stability}(b)), and they develop into linked half-quantum
vortex rings, i.e., a skyrmion.
This result implies that a U(1) vortex ring as in Fig.~\ref{f:stability}(a)
is dynamically unstable against splitting into two half-quantum vortex rings.
Such dynamical instabilities in two-component vortices merit further study.

\subsection{Trapped system}
\label{s:trap}

We consider a realistic system, in which a BEC of $^{87}{\rm Rb}$ atoms is
confined in a harmonic potential.
For simplicity, we assume that the two components feel the same isotropic
harmonic potential given by $V_{\rm trap}(\bm{r}) = m \omega^2 (x^2 + y^2 +
z^2) / 2$ with a trap frequency $\omega = 2\pi \times 100$ Hz.
For components 1 and 2, we assume the hyperfine states $|F = 1, m_F = 1
\rangle$ and $|F = 1, m_F = 0 \rangle$ of an $^{87}{\rm Rb}$ atom, where $F$
is the hyperfine spin and $m_F$ is the magnetic sublevel.
The $s$-wave scattering lengths are therefore $a_{11} = a_{12} = 100.4 a_B$
and $a_{22} = 100.86 a_B$ with $a_B$ being the Bohr
radius~\cite{Kempen,Widera}, and the miscible condition $a_{11} a_{22} >
a_{12}^2$ is satisfied.
The numbers of atoms are $N_1 = N_2 = 3 \times 10^5$.
In this subsection, we normalize the length, time, and density by
$a_{\rm ho} = [\hbar / (m \omega)]^{1/2}$, $\omega^{-1}$, and
$N_j / a_{\rm ho}^3$, respectively. 

\begin{figure}[tb]
\includegraphics[width=8.0cm]{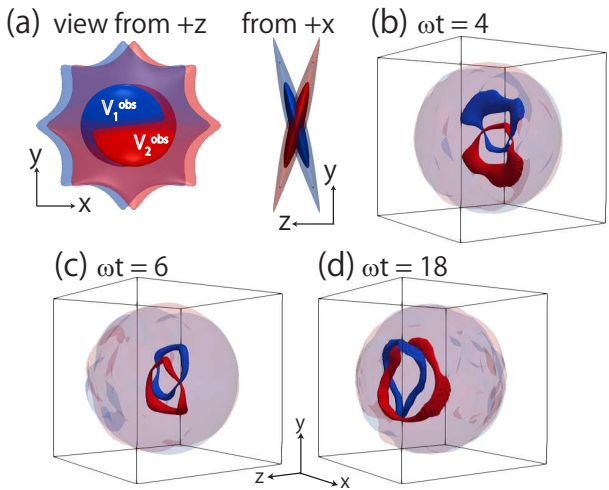}
\caption{
  (a) Isosurfaces of the obstacle potentials $V_1^{\rm obs} = V_2^{\rm obs}
  = 2 V_0$ (inner) and $1.05 V_0$ (outer) in Eqs.~(\ref{V1})-(\ref{V12})
  with $R = 3 a_{\rm ho}$, $\lambda = \pi / 12$, $\delta = 0.5 a_{\rm ho}$,
  $b_1 = b_2 = 0.85$, and $V_0 = 10 \hbar \omega$
  (b)-(d) Dynamics of skyrmion generation, where $N_1 = N_2 = 3 \times 10^5$
  atoms are confined in an isotropic harmonic potential with a frequency
  $\omega = 2\pi \times 100$ Hz.
  The obstacle potentials in (a) are moved at a velocity
  $v_z = 0.5 a_{\rm ho} \omega$, and the strength $V_0$ is linearly ramped
  down and vanishes at $\omega t = 3$.
  The isodensity surfaces of both components ($|\psi_1|^2 = |\psi_2|^2 =
  10^{-4} N_j / a_{\rm ho}^3$) are shown, where the surfaces at
  $r > 6 a_{\rm ho}$ are made transparent.
  The size of the boxes in (b)-(d) is $(17.92 a_{\rm ho})^3$.
  See the Supplemental Material for a video of the dynamics in
  (b)-(d)~\cite{movies}.
}
\label{f:trap}
\end{figure}
A three-dimensional isolated Gaussian potential as in Eq.~(\ref{Vj}) is
difficult to realize in experiments if laser beams penetrating through
the BEC are used to produce the potential.
Instead, we use a potential produced by four Gaussian laser beams crossed
with an angle $\pi / 4$ as
\begin{eqnarray} \label{V1}
  f_{\rm cross}(\bm{r}) & = & \sum_{n=0}^3
  e^{-[(x \cos\frac{n\pi}{4} - y \sin\frac{n\pi}{4})^2 + 4 z^2] / R^2}.
\end{eqnarray}
Using these crossed Gaussian beams, we can increase the peak height of the
potential, suppressing the effect of the incoming and outgoing beams,
which mimics an isolated potential as in Eq.~(\ref{Vj}).
We tilt the potential in Eq.~(\ref{V1}) about the $x$ axis by an angle
$\pm \lambda$ and shift it in the $x$ direction by $\pm\delta$, giving
\begin{eqnarray}
  f_\pm(\bm{r}) & = & f_{\rm cross}(x \pm \delta, \eta, \zeta),
  \nonumber \\
  \eta & = & y \cos\lambda \mp z \sin\lambda, \nonumber \\
  \zeta & = & z \cos\lambda \pm y \sin\lambda.
\end{eqnarray}
In general, using near-resonant laser beams, the fields $f_\pm(\bm{r})$
generated by laser frequencies $\omega_\pm$ produce different potentials for
components 1 and 2~\cite{Kim, Kim2}, and we assume an obstacle potential of
the form,
\begin{eqnarray} \label{V12}
  V_1^{\rm obs}(\bm{r}) & = & V_0 [f_+(\bm{r}) + b_1 f_-(\bm{r})],
  \nonumber \\
  V_2^{\rm obs}(\bm{r}) & = & V_0 [f_-(\bm{r}) + b_2 f_+(\bm{r})],
\end{eqnarray}
where $V_0$, $b_1$, and $b_2$ are constants.
Figure~\ref{f:trap}(a) shows the isosurfaces of the potentials in
Eq.~(\ref{V12}) with $R = 3 a_{\rm ho}$, $\lambda = \pi / 12$, $\delta =
0.5 a_{\rm ho}$, $b_1 = b_2 = 0.85$, and $V_0 = 10 \hbar \omega$.
Although the outer surfaces ($V_1^{\rm obs} = V_2^{\rm obs} = 1.05 V_0$)
are star-shaped reflecting the superposition of the four Gaussian beams, the
inner surfaces ($V_1^{\rm obs} = V_2^{\rm obs} = 2 V_0$) have oblate shapes,
similar to those of the oblate potentials in Fig.~\ref{f:single}(a).
Replacing $z$ with $z(t) = z - z_0 - v_z t$ in Eqs.~(\ref{V1})-(\ref{V12}),
the obstacle potential can be moved in the $z$ direction at a velocity
$v_z$.
The total potential in the GP equation thus has the form, $V_j(\bm{r}, t) =
V_j^{\rm obs}(\bm{r}, t) + V_{\rm trap}(\bm{r})$.

Figures~\ref{f:trap}(b)-\ref{f:trap}(d) show the dynamics of skyrmion
generation (see also a video in the Supplemental Material~\cite{movies}),
where the initial state is the ground state with the obstacle potential in
Fig.~\ref{f:trap}(a) at $z_0 = -2$.
In the time evolution, the obstacle potential is moved at a velocity $v_z =
0.5 a_{\rm ho} \omega$ and at the same time the magnitude $V_0$ is linearly
ramped down and vanishes at $\omega t = 3$.
We can see that vortex rings are generated in both components and they are
linked with each other, forming a skyrmion.
The skyrmion moves in the $+z$ direction, and when it reaches the edge
of the BEC, the vortex rings expand and move back in the $-z$ direction
along the periphery of the BEC.
The skyrmion is then destroyed by the excitations left in the BEC.

\section{Conclusions}
\label{s:conc}

We proposed a method to generate three-dimensional skyrmions behind an
obstacle moving in a two-component BEC.
The linked-vortex configuration of a skyrmion in Fig.~\ref{f:skyrmion}(d) is
more suitable than that in Fig.~\ref{f:skyrmion}(a) for hydrodynamic
generation.
We proposed a shape of an obstacle potential in which two oblates bite into
each other, as shown in Fig.~\ref{f:single}(a), and numerically demonstrated
that it can create a skyrmion, as shown in
Figs.~\ref{f:single}(b)-\ref{f:single}(g).
We also showed that skyrmions are released behind the obstacle successively,
and these skyrmions can be stabilized by a guiding potential
(Fig.~\ref{f:multi}).
Such skyrmion generation can be realized in a realistic experimental system
with a feasible number of atoms ($6 \times 10^5$ atoms of $^{87}{\rm Rb}$),
as shown in Fig.~\ref{f:trap}.

The potential used in Fig.~\ref{f:trap} needs eight Gaussian laser beams,
which requires considerable experimental effort.
If we optimize the parameters, it may be possible to reduce the number of
laser beams in experiments.
Considering that a vortex ring in a single-component BEC can be generated by
a single Gaussian laser beam~\cite{Saito}, it may be possible to generate a
skyrmion by only two laser beams.
Machine-learning techniques will be useful in determining an optimized
protocol~\cite{Saito}.
Also, through machine-learning optimization, we expect that the
lifetime of skyrmions after generation can be prolonged by optimizing
the potential shapes, the manner of moving the potentials, and other
parameters.
If successive long-lifetime skyrmions can be generated and if some
symmetry breaking instability arises in the flow near the obstacle,
alternate generation of skyrmions and antiskyrmions might occur, resulting
in a B\'enard-von K\'arm\'an-like skyrmion street.

\begin{acknowledgments}
We wish to thank A. Yamamoto for contributing to the early stage of this
work.
This work was supported by JSPS KAKENHI Grant Number JP20K03804.
\end{acknowledgments}


\begin{thebibliography}{99}

\bibitem{Skyrme}
  T. H. R. Skyrme, Proc. R. Soc. London A {\bf 260}, 127 (1961);
  Nucl. Phys. {\bf 31}, 556 (1962).

\bibitem{Lee}
  D.-H. Lee and C. L. Kane,
  Phys. Rev. Lett. {\bf 64}, 1313 (1990).

\bibitem{Brey}
  L. Brey, H. A. Fertig, R. C\^ot\'e, and A. H. MacDonald,
  Phys. Rev. Lett. {\bf 75}, 2562 (1995).

\bibitem{Schmeller}
  A. Schmeller, J. P. Eisenstein, L. N. Pfeiffer, and K. W. West,
  Phys. Rev. Lett. {\bf 75}, 4290 (1995).

\bibitem{Bogdanov}
  A. N. Bogdanov and U. K. R\"o\ss ler,
  Phys. Rev. Lett. {\bf 87}, 037203 (2001);
  U. K. R\"o\ss ler, A. N. Bogdanov, and C. Pfleiderer,
  Nature (London) {\bf 442}, 797 (2006).

\bibitem{Bogdanov2}
  A. N. Bogdanov, U. K. R\"o\ss ler, and A. A. Shestakov,
  Phys. Rev. E {\bf 67}, 016602 (2003).

\bibitem{Ruo}
  J. Ruostekoski and J. R. Anglin,
  Phys. Rev. Lett. {\bf 86}, 3934 (2001).

\bibitem{Khawaja}
  U. Al Khawaja and H. Stoof,
  Nature (London) {\bf 411}, 918 (2001).

\bibitem{Yu}
  X. Z. Yu, Y. Onose, N. Kanazawa, J. H. Park, J. H. Han, Y. Matsui,
  N. Nagaosa, and Y. Tokura,
  Nature (London) {\bf 465}, 901 (2010).

\bibitem{Romming}
  N. Romming, C. Hanneken, M. Menzel, J. E. Bickel, B. Wolter, K. von
  Bergmann, A. Kubetzka, and R. Wiesendanger,
  Science {\bf 341}, 636 (2013).

\bibitem{Leslie}
  L. S. Leslie, A. Hansen, K. C. Wright, B. M. Deutsch, and N. P. Bigelow,
  Phys. Rev. Lett. {\bf 103}, 250401 (2009).

\bibitem{Choi}
  J.-Y. Choi, W. J. Kwon, and Y. Shin,
  Phys. Rev. Lett. {\bf 108}, 035301 (2012).

\bibitem{Lee2}
  W. Lee, A. H. Gheorghe, K. Tiurev, T. Ollikainen, M. M\"ott\"onen, and
  D. S. Hall,
  Sci. Adv. {\bf 4}, 3820 (2018).

\bibitem{Tiurev}
  K. Tiurev, T. Ollikainen, P. Kuopanportti, M. Nakahara, D. S. Hall, and
  M. M\"ott\"onen,
  New J. Phys. {\bf 20}, 055011 (2018).

\bibitem{Zamora}
  R. Zamora-Zamora and V. Romero-Rochin,
  J. Phys. B: At. Mol. Opt. Phys. {\bf 51}, 045301 (2018).

\bibitem{Luo}
  H.-B. Luo, L. Li, and W.-M. Liu,
  Sci. Rep. {\bf 9}, 18804 (2019).

\bibitem{Sasaki}
  K. Sasaki, N. Suzuki, and H. Saito,
  Phys. Rev. A {\bf 83}, 053606 (2011).

\bibitem{Kawakami}
  T. Kawakami, T. Mizushima, M. Nitta, and K. Machida,
  Phys. Rev. Lett. {\bf 109}, 015301 (2012).

\bibitem{Zhang}
  G. Chen, T. Li, and Y. Zhang,
  Phys. Rev. A {\bf 91}, 053624 (2015).

\bibitem{Nitta}
  M. Nitta, K. Kasamatsu, M. Tsubota, and H. Takeuchi,
  Phys. Rev. A {\bf 85}, 053639 (2012).

\bibitem{Parmee}
  C. D. Parmee, M. R. Dennis, and J. Ruostekoski,
  arXiv:2109.13927.

\bibitem{Khawaja2}
  U. Al Khawaja and H. T. C. Stoof
  Phys. Rev. A {\bf 64}, 043612 (2001).

\bibitem{Battye}
  R. A. Battye, N. R. Cooper, and P. M. Sutcliffe,
  Phys. Rev. Lett. {\bf 88}, 080401 (2002).

\bibitem{Zhang2}
  Y. Zhang, W.-D. Li, L. Li, and H. J. W. M\"uller-Kirsten,
  Phys. Rev. A {\bf 66}, 043622 (2002).

\bibitem{Zhai}
  H. Zhai, W. Q. Chen, Z. Xu, and L. Chang,
  Phys. Rev. A {\bf 68}, 043602 (2003).

\bibitem{Savage}
  C. M. Savage and J. Ruostekoski,
  Phys. Rev. Lett. {\bf 91}, 010403 (2003).

\bibitem{Ruo2}
  J. Ruostekoski,
  Phys. Rev. A {\bf 70}, 041601(R) (2004).

\bibitem{Wuster}
  S. W\"uster, T. E. Argue, and C. M. Savage,
  Phys. Rev. A {\bf 72}, 043616 (2005).

\bibitem{Herbut}
  I. F. Herbut and M. Oshikawa,
  Phys. Rev. Lett. {\bf 97}, 080403 (2006).

\bibitem{Tokuno}
  A. Tokuno, Y. Mitamura, M. Oshikawa, and I. F. Herbut,
  Phys. Rev. A {\bf 79}, 053626 (2009).

\bibitem{Price}
  H. M. Price and N. R. Cooper,
  Phys. Rev. A {\bf 83}, 061605(R) (2011).

\bibitem{Kaneda}
  T. Kaneda and H. Saito,
  Phys. Rev. A {\bf 93}, 033611 (2016).

\bibitem{Cooper}
  R. G. Cooper, M. Mesgarnezhad, A. W. Baggaley, and C. F. Barenghi,
  Sci. Rep. {\bf 9}, 10545 (2019).

\bibitem{Kleckner}
  D. Kleckner and W. T. M. Irvine,
  Nat. Phys. {\bf 9}, 253 (2013).

\bibitem{Frisch}
  T. Frisch, Y. Pomeau, and S. Rica,
  Phys. Rev. Lett. {\bf 69}, 1644 (1992).

\bibitem{Jackson}
  B. Jackson, J. F. McCann, and C. S. Adams,
  Phys. Rev. Lett. {\bf 80}, 3903 (1998).

\bibitem{Nore}
  C. Nore, C. Huepe, and M. E. Brachet,
  Phys. Rev. Lett. {\bf 84}, 2191 (2000).

\bibitem{Inouye}
  S. Inouye, S. Gupta, T. Rosenband, A. P. Chikkatur, A. G\"orlitz,
  T. L. Gustavson, A. E. Leanhardt, D. E. Pritchard, and W. Ketterle,
  Phys. Rev. Lett. {\bf 87}, 080402 (2001).
  
\bibitem{Neely}
  T. W. Neely, E. C. Samson, A. S. Bradley, M. J. Davis, and B. P. Anderson,
  Phys. Rev. Lett. {\bf 104}, 160401 (2010).

\bibitem{Kwon}
  W. J. Kwon, G. Moon, S. W. Seo, and Y. Shin,
  Phys. Rev. A {\bf 91}, 053615 (2015).

\bibitem{Kwon2}
  W. J. Kwon, S. W. Seo, and Y. Shin,
  Phys. Rev. A {\bf 92}, 033613 (2015).

\bibitem{SasakiL}
  K. Sasaki, N. Suzuki, and H. Saito,
  Phys. Rev. Lett. {\bf 104}, 150404 (2010).

\bibitem{Kwon3}
  W. J. Kwon, J. H. Kim, S. W. Seo, and Y. Shin,
  Phys. Rev. Lett. {\bf 117}, 245301 (2016).

\bibitem{Seo}
  S. W. Seo, S. Kang, W. J. Kwon, and Y. Shin,
  Phys. Rev. Lett. {\bf 115}, 015301 (2015).

\bibitem{Seo2}
  S. W. Seo, W. J. Kwon, S. Kang, and Y. Shin,
  Phys. Rev. Lett. {\bf 116}, 185301 (2016).

\bibitem{Nardin}
  G. Nardin, G. Grosso, Y. L\'eger, B. Pietka, F. Morier-Genoud, and
  B. Deveaud-Pl\'edran,
  Nat. Phys. {\bf 7}, 635 (2011).

\bibitem{recipe}
W. H. Press, S. A. Teukolsky, W. T. Vetterling, and B. P. Flannery,
{\it Numerical Recieps}, 3rd ed. (Cambridge Univ. Press, Cambridge, 2007).  

\bibitem{movies}
See Supplemental Material at http://link.aps.org/supplemental/... for videos
of the dynamics.

\bibitem{Gudnason}
  S. B. Gudnason and M. Nitta,
  Phys. Rev. D {\bf 101}, 065011 (2020).

\bibitem{Kim}
  J. H. Kim, D. Hong, and Y. Shin,
  Phys. Rev. A {\bf 101}, 061601(R) (2020).

\bibitem{Kim2}
  J. H. Kim, D. Hong, K. Lee, and Y. Shin,
  Phys. Rev. Lett. {\bf 127}, 095302 (2021).

\bibitem{Kempen}
  E. G. M. van Kempen, S. J. J. M. F. Kokkelmans, D. J. Heinzen, and
  B. J. Verhaar,
  Phys. Rev. Lett. {\bf 88}, 093201 (2002).

\bibitem{Widera}
  A. Widera, F. Gerbier, S. F\"olling, T. Gericke, O. Mandel, and I. Bloch,
  New J. Phys. {\bf 8}, 152 (2006).

\bibitem{Saito}
  H. Saito,
  J. Phys. Soc. Jpn. {\bf 89}, 074006 (2020).


\end{thebibliography}
\end{document}